\documentclass[pdflatex,sn-mathphys-num]{sn-jnl}
\usepackage{graphicx}%
\usepackage{multirow}%
\usepackage{amsmath,amssymb,amsfonts}%
\usepackage{amsthm}%
\usepackage{mathrsfs}%
\usepackage[title]{appendix}%
\usepackage{xcolor}%
\usepackage{textcomp}%
\usepackage{manyfoot}%
\usepackage{booktabs}%
\usepackage{algorithm}%
\usepackage{algorithmicx}%
\usepackage{algpseudocode}%
\usepackage{listings}%
\usepackage{siunitx}
\usepackage{todonotes}
\usepackage{multirow}
\usepackage{tablefootnote}
\begin{document}

\title[Article Title]{Cooperative-Memory Photonic Reservoir using Modulation Nonlinearity: Circumventing the Speed Constraints of Nonlinear Silicon Microring Resonators}

\author*[1]{\fnm{Amideddin} \sur{Mataji-Kojouri}}\email{amideddin.mataji\_kojouri@tu-dresden.de}

\author[2]{\fnm{Sebastian} \sur{Kühl}}\email{sk@tf.uni-kiel.de}

\author[2]{\fnm{Mohammad} \sur{Seifi Laleh}}\email{mola@tf.uni-kiel.de}
\author[2]{\fnm{Stephan} \sur{Pachnicke}}\email{stephan.pachnicke@tf.uni-kiel.de}
\author*[1]{\fnm{Kambiz} \sur{Jamshidi}}\email{kambiz.jamshidi@tu-dresden.de}

\affil*[1]{\orgdiv{Integrated Photonic Devices Group, Chair of RF and Photonics Engineering}, \orgname{TU  Dresden}, \orgaddress{\street{Helmholzstr. 18}, \city{Dresden}, \postcode{01069}, \country{Germany}}}

\affil[2]{\orgdiv{Chair of Communications}, \orgname{Kiel University}, \orgaddress{\street{Kaiserstr. 2}, \city{Kiel}, \postcode{24143}, \country{Germany}}}

\abstract{
Complex dynamics of silicon microring resonators loaded by delayed feedback elements enable high-speed photonic reservoir computing. Implementing feedback is especially challenging when the required delay should match the time scales of silicon's nonlinearities. To increase the computation speed and preclude any need for very long delay lines, we avoid relying on silicon's nonlinearity and merely employ either amplitude or phase modulation along with direct detection.
By supplementing its memory with that of the electronic output layer, the proposed photonic reservoir composed of a Mach-Zehnder interferometer and a microring resonator is predicted to perform computations one order of magnitude faster than those based on silicon's nonlinearity with its speed only limited by the modulation/detection bandwidth. This reservoir performs accurately in NARMA-10, Mackey-Glass, and Santa-Fe prediction tasks, and enables signal equalization in optical communication systems.   
}

\maketitle

\section{Introduction}\label{sec1}

Photonic reservoir computers employ complex dynamics of a photonic system to process data and perform computational tasks at speeds readily exceeding 1 GHz \cite{van2017advances, gauthier2021next,brunner2019photonic}. The reservoir is a randomly connected network with static weights that is fed by an external signal at its input layer. The output layer is trained to produce desired output by reading the state of the reservoir. Computation with these systems is enabled by their complex dynamics that can be described by nonlinear delay differential equations \cite{appeltant2011information,ikeda1987high}. 
Due to their rich nonlinear dynamics \cite{shetewy2024demonstration, gray2020thermo} and relatively simple implementation, silicon microring resonators (MRRs) have been widely investigated for implementing photonic reservoir computers \cite{li2021micro, borghi2021reservoir, donati2021microring, donati2024time,giron2024effects, ren2024photonic} 
Nonlinear phenomena in a silicon MRR include two-photon absorption (TPA), which generates free carriers that subsequently contribute to free-carrier absorption (FCA) and free-carrier dispersion (FCD), as well as thermo-optic (TO) effect and the intrinsic optical Kerr effect of silicon. These nonlinear optical phenomena occur in different time scales. The fastest is optical Kerr effect which is almost instantaneous due to its electronic origin. FCD and FCA exhibit intermediate response times on the order of a few nanoseconds determined by the carrier lifetime. The TO effect demonstrates the slowest response dominated by the heat diffusion processes, with characteristic times ranging from tens to hundreds of nanoseconds \cite{chen2012bistability,johnson2006self}. Although the Kerr effect is present in silicon MRRs, it is usually overshadowed by the relatively slower but significantly stronger FCD/FCA effects which are the dominant nonlinear mechanisms in a silicon MRR. To obtain a faster response, it is possible to decrease the free carrier lifetime by applying a reverse bias, which facilitates carrier extraction. However, faster decay also weakens the FCD/FCA processes and increases their corresponding threshold power, thus limiting the performance of the reservoir.
silicon MRRs can be also loaded by delayed feedback to exhibit a better performance in computing compared to a single MRR. To exploit silicon's FCD/FCA nonlinearity in these structures, implementing a complex feedback element with a time delay matching the free carrier lifetime (a few nanoseconds) is inevitable \cite{donati2021microring,donati2024time,ren2024photonic}. 
In order to avoid these challenges, and the fact that the FCD/FCA nonlinearity does not efficiently improve the performance of the computation \cite{donati2021microring,donati2024time,ren2024photonic}, here we restrict our study to the linear regime. 
Another important factor influencing computation speed is the masking process where the input sequence is multiplied element-wise by another sequence called mask to increase the input dimensionality. With a constant modulation speed, the computation speed is inversely proportional to the mask length. We find that when keeping the complexity of the electronic output layer identical, the reservoir computer with a shorter mask can also produce a more accurate output because the photonic and electronic layers cooperatively contribute to the memory of the reservoir network.   

In light of these findings, we propose a photonic reservoir composed of an unbalanced Mach-Zehnder interferometer (MZI) along with an all-pass MRR. To increase the computation speed we restrict the ring to the linear regime and use a short mask sequence. 
The performance of this MZI-MRR-based reservoir is numerically investigated in time-series prediction tasks. Without adding any extra complexity, the proposed reservoir can predict time series at speeds only limited by the modulation speed. We also show that it can be employed for signal equalization in a short-reach fiber optic transmission system. 

The structure of the paper is as follows. In Section \ref{sec2} we summarize the linear and nonlinear models used to describe the behavior of the MRRs. In Section \ref{sec3}, we first investigate the impact of nonlinear optical phenomena on the performance of an MRR-based reservoir. Then we study the impact of mask length and output layer dimensionality on the performance of the photonic reservoir. In Section \ref{sec4} we discuss the performance of the MZI-MRR structure in time series and signal equalization tasks.

\section{Modelling silicon microring resonators}\label{sec2}
In this section, we discuss the linear and nonlinear models for predicting the response of an all-pass silicon MRR that is excited through a loss-less coupler. These models can be easily extended to more complicated configurations as will be briefly discussed later. 
In linear regime, we can use a transfer matrix formalism to find the transfer function for the photonic integrated circuit in the frequency domain. For an all-pass ring resonator, the transfer function is
\begin{align}
T(\omega)=\frac{{\sigma-e^{-\gamma\left(\omega\right) l}}}{{1-\sigma e^{-\gamma\left(\omega\right) l}}}, \label{eq:Transferfunction}
\end{align}
where $l$ is the circumference of the ring resonator, $\gamma\left(\omega\right)=\alpha+jk(\omega)$ is the complex propagation constant, and $\sigma$ is the self-coupling coefficient of the lossless coupler. The frequency dependency of the propagation constant and the coupling coefficients can be calculated using a full-wave modal solver and introduced to the linear model. To calculate the output of the ring resonator excited by an input signal in time domain, one should employ the fast Fourier transform (FFT). 
To approximate the behavior of the ring resonator in the nonlinear regime, we employ a coupled-mode theory (CMT) analysis \cite{haus1984waves,zhang2013multibistability,gray2020thermo,chen2012bistability}. When a silicon ring resonator is excited by an optical input in the near-infrared range, large circulating power in the ring resonator, which is a consequence of the field enhancement at resonance, gives rise to TPA which in turn produces free carriers in the silicon waveguide. The presence of free carriers in silicon increases the loss due to FCA and changes its refractive index due to FCD. Absorption of optical power in the ring can increase the temperature of the resonator and change its refractive index due to the TO effect. Finally, third-order nonlinear susceptibility of silicon (optical Kerr effect) alters the effective index of the waveguide proportional to the circulating light power. The following CMT equations approximate the dynamics of the complex amplitude of the mode in the MRR $(u)$, electron-hole pair density $(N)$, and temperature change $(\Delta T)$ for an all-pass silicon ring resonator \cite{johnson2006self,zhang2013multibistability,chen2012bistability}: 
\begin{align}
\frac{d u}{d t} = &\ \left\{\vphantom{\left[\overbrace{\frac{n_2 c_0}{n_0 V_{\mathrm{Kerr}}} \left|u \right|^2}^{\mathrm{Kerr}}\right]} \right. \nonumber
\frac{-j\omega_\mathrm{L}}{n_0} \left[\overbrace{\frac{n_2 c_0}{n_0 V_{\mathrm{Kerr}}} \left|u \right|^2}^{\mathrm{Kerr}} - \overbrace{\left( \sigma_{\mathrm{r_1}}N + \sigma_{\mathrm{r_2}}N^{0.8} \right)}^{\mathrm{FCD}} + \overbrace{\kappa_\mathrm{\theta} \Delta T}^{\mathrm{TO}} \right] \\ \nonumber
&+ j\overbrace{\left(\omega_0 - \omega_\mathrm{L} \right)}^{\mathrm{detuning}} - \frac{c_0}{2n_0}\left[\overbrace{\alpha}^{\mathrm{LL}} + \overbrace{\frac{\beta_2 c_0}{n_0V_{\mathrm{TPA}}} \left|u\right|^2}^{\mathrm{TPA}} + \overbrace{\sigma_{\mathrm{FCA}}N}^{\mathrm{FCA}} \right] \left. \vphantom{\left[\overbrace{\frac{n_2 c_0}{n_0 V_{\mathrm{Kerr}}} \left|u \right|^2}^{\mathrm{Kerr}}\right]} \right\} u \\
&+ \sqrt{\Gamma_\mathrm{c}} \, s_{\mathrm{in}}\textcolor{black}{\left(t\right)}, \label{eq:CMT mode amplitude}
\end{align}
\begin{align}
\frac{d N}{d t} = &\ \frac{c_0^2\beta_2}{n_0^2 2\hbar\omega_\mathrm{L} V_{\mathrm{FCA}}^2}\left|u\right|^4- \frac{N}{\tau_{\mathrm{car}}},\label{eq:CMT carrier density}
\end{align}
\begin{align}
\frac{d \Delta T}{d t} = &\ \frac{\left|u \right|^2}{\rho_{\mathrm{Si}} c_{\mathrm{Si}}V_{\mathrm{eff}}}\left(\frac{\textcolor{black}{\alpha_{\mathrm{abs}}} c_0}{n_0}+\frac{c_0^2\beta_2\left|u \right|^2}{n_0^2V_{\mathrm{TPA}}} +\frac{\sigma_{\mathrm{FCA}} N c_0}{n_0}\right)-\frac{\Delta T}{\tau_{\mathrm{th}}}. \label{eq:CMT Temperature}
\end{align}
Here, $U=\left| u\right|^2$ is the mode energy, $\omega_L$ is the angular frequency of the input laser light, $\omega_0$ is the resonance angular frequency, $c_0$ is the speed of light in vacuum, $n_0$ \textcolor{black}{is the group index of the waveguide}, ${n_2 c_0}\left|u \right|^2/{(n_0 V_{\mathrm{Kerr}})}$ is the index change due to the Kerr effect with $n_2$ the Kerr coefficient, $V_{\mathrm{Kerr}}=A_{\mathrm{Kerr}}L$ the Kerr nonlinear volume, $A_{\mathrm{Kerr}}$ the effective cross-section of the waveguide for the Kerr effect, and $L$ the circumference of the MRR. The index change due to the FCD effect is introduced to the equations by the $ \sigma_{\mathrm{r_1}}N + \sigma_{\mathrm{r_2}}N^{0.8} $ term where $\sigma_{\mathrm{r_1}}=8.8\times10^{-28} \, m^3$ and $\sigma_{\mathrm{r_2}}=1.35\times10^{-22} \,m^{2.4}$ relate the refractive index change to the electron and hole densities, respectively \cite{soref1987electrooptical}. The index change due to the thermo-optic effect is given by $k_{\theta} \Delta T$ with $k_{\theta}$ the TO coefficient of silicon, and $\Delta T$ the temperature change of the ring resonator. The total linear loss (LL) of the MRR is $\alpha=\alpha_{ring}+\alpha_{c}$ where $\alpha_{ring}=\alpha_{abs}+\alpha_{rad}+\alpha_{sca}$ is the linear loss of the ring composed of three parts with $\alpha_{abs}$ the linear absorption loss which contributes to the heat generation and temperature increase in the silicon MRR, $\alpha_{rad}$ the radiation loss due to the coupling of light to the higher-order modes of the waveguide, and $\alpha_{sca}$ the scattering loss. $\alpha_{c}$ is the coupling loss. The nonlinear loss due to the TPA is $\beta_2 c_0 \left|u\right|^2/({n_0 V_{\mathrm{TPA}}})$ with $\beta_2$ the TPA coefficient and $V_{\mathrm{TPA}}=V_{\mathrm{Kerr}}$ the TPA volume. The FCA loss is ${\sigma_{\mathrm{FCA}}N}$ with $\sigma_{\mathrm{FCA}}$ the FCA coefficient. \textcolor{black}{$\Gamma_c$ is the coupling coefficient between the straight waveguide and the ring waveguide given by $\Gamma_c=c\alpha_c/n_0$ and $s_{\mathrm{in}}(t)$ is the power wave amplitude for the incident light in the waveguide so that the $|s_{\mathrm{in}}(t)|^2$ gives the power carried by the incident light.} In the \eqref{eq:CMT mode amplitude}, Kerr effect, FCD, and TO effect contribute to the resonance shift, and the linear loss, TPA, and FCA affect the decay rate of the complex mode amplitude. $V_{\mathrm{FCA}}=A_{\mathrm{FCA}}L$ is the FCA volume, with $A_{\mathrm{FCA}}$ the effective cross section of the FCA effect, $\hbar$ is the reduced Planck's constant, and $\tau_{\mathrm{car}}$ is the free carrier lifetime.  In \eqref{eq:CMT carrier density}, free carrier's generation is related to the TPA effect and their recombination lifetime is assumed constant ($\tau_{\mathrm{car}}$). $\rho_{\mathrm{Si}}$ and $c_{\mathrm{Si}}$ are the density and the specific heat capacity of silicon, respectively, $V_{\mathrm{eff}}=A_{\mathrm{eff}}L$ is the effective volume of the silicon waveguide with $A_{\mathrm{eff}}$ the effective cross-section of the waveguide \cite{zhang2013multibistability}, and $\tau_{\mathrm{th}}$ is the thermal decay time. In \eqref{eq:CMT Temperature}, the absorption part of the linear loss along with the TPA, and FCA nonlinear losses are the three sources for temperature increase and we assume that the temperature decays with a single decay time for simplicity \cite{borghi2021modeling}. Given the large differences in the magnitudes of the complex mode amplitude, free carrier density, and temperature change, we employ a normalized set of equations that uses dimensionless parameters to facilitate the numerical analysis \cite{zhang2013multibistability}. The output power is $|s_{\mathrm{out}}(t)|^2$ with $s_{\mathrm{out}}(t)$ the complex amplitude of the output light $s_{\mathrm{out}}(t)=\sqrt{P_\mathrm{in}}-\sqrt{\Gamma_\mathrm{c}} u(t)$. 

In our calculations, we assume a group index of \textcolor{black}{$n_0$= 4.1} and a loss of 2 dB/cm for the silicon waveguides. Values for the other parameters of the nonlinear model are summarized in \textcolor{black}{Table S1}.

\begin{figure}
\centering
\includegraphics[width=0.95\columnwidth]{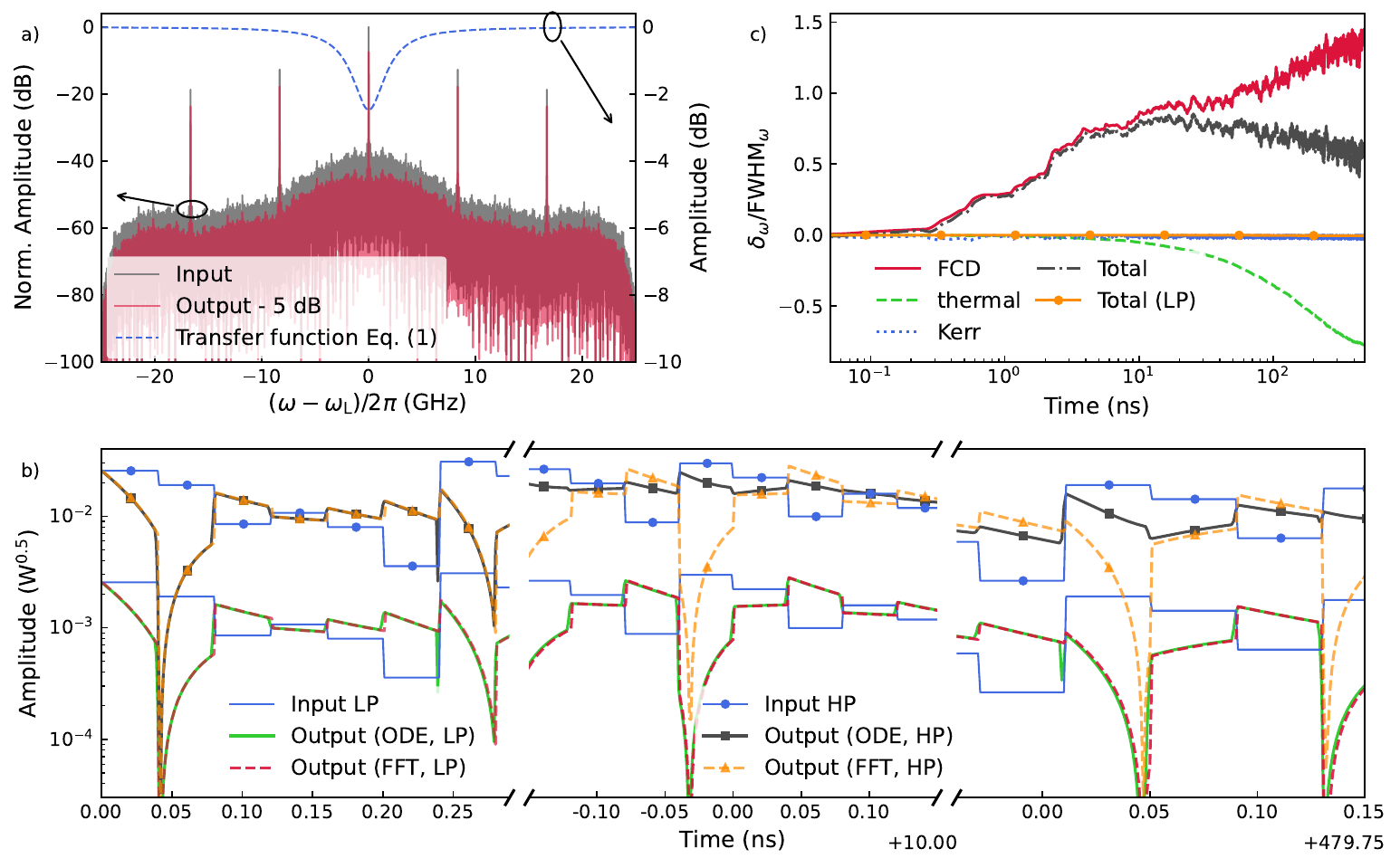}
\caption{\label{fig:FIG1} Linear and nonlinear response of an all-pass ring resonator. a) Frequency spectra of an AM-modulated signal masked by a 3-element mask sequence (gray), the transfer function of a typical all-pass ring resonator (blue, left axis), and the output spectra (red). b) Comparing the time response of the ring resonator for the LP and HP case using both linear and nonlinear models. In the LP regime, the models' predictions are in good agreement. c) Normalized frequency detuning vs. time, showing different time-scales and the strength for FCD/FCA, TO, and Kerr effect when the ring is excited by a relatively high-power (HP) optical input signal ($P_{in}=1\,\textrm{mW}$). For the low-power (LP) case ($P_{in}=10\,\mu\textrm{W}$), the total frequency detuning is always negligible. Inset shows a schematic of an all-pass ring resonator coupled to a nearby waveguide.}
\end{figure}

We now compare the linear solution with results from the nonlinear model in a silicon ring resonator to understand the input power levels for which the linear model correctly predicts the system's response. In Fig. \ref{fig:FIG1}, the response of the linear and nonlinear model for an over-coupled all-pass ring resonator with a radius of 10 $\mu$m and a loaded quality factor of $Q_{l}=8.15\times10^4$ are presented. Here, the laser is centered around $\lambda_\mathrm{L}=1550\, \mathrm{nm}$ and we assume a normalized detuning of $\delta_\omega/\mathrm{FWHM}_\omega=0$ with FWHM$_\omega$ the full width at half maximum of the resonance. To resemble a typical computing scenario, the amplitude of the laser light is modulated using a sample-and-hold scheme with a modulation index of 1. The input is a pseudo-random sequence of length 4000 with a uniform distribution in the range between 0 and 1 which is multiplied element-wise by a periodic mask of length three (M=[0.9, 0.5, 0.1]). Pulse duration ($\tau_p$) is $\tau_p=40\, $ps. The maximum instantaneous power of light in the bus waveguide is either 1 mW (high power (HP)) or 10 $\mu$W (low power (LP)). In Fig. \ref{fig:FIG1}a, the normalized spectrum of the masked input light and the output light is demonstrated (left axis) around the center frequency of the laser. The amplitude of the transfer function of the all-pass MRR is also shown in this figure (right axis). The spectra exhibit peaks at the harmonics of the fundamental frequency of the masking signal, and its bandwidth is 25 GHz. Considering the initial detuning of zero, the signal spectrum is aligned with the resonance of the transfer function. Figure \ref{fig:FIG1}b shows the input and the calculated output signals in the time domain achieved using either the linear or the nonlinear model at three different periods of time for either the LP or the HP cases. For the LP case, both models are always in agreement. For the HP case, the models are only initially in agreement, but after passing some time their predictions for the output of the ring are very different. Figure \ref{fig:FIG1}c demonstrates the detuning of the ring resonator due to the nonlinear effects in the MRR in both LP and HP cases. For the LP case, the total detuning is always negligible, confirming that a linear model suffices for predicting the MRR's response. For the HP case, the detuning is significant mostly due to the FCD/FCA and TO effects. Figure \ref{fig:FIG1}c also illustrates different time scales of these nonlinear effects for typical MRRs which are several nanoseconds for the FCD/FCA and tens of nanoseconds for the TO effect. Although the Kerr effect is fast, it is not strong enough to significantly impact the response and is usually overshadowed by the slow FCD/FCA effect.

\section{Reservoir Computing with a Linear PIC}\label{sec3}
\subsection{All-pass ring resonator}\label{RingConfig}
\begin{figure}
\centering
\includegraphics[width=0.95\columnwidth]{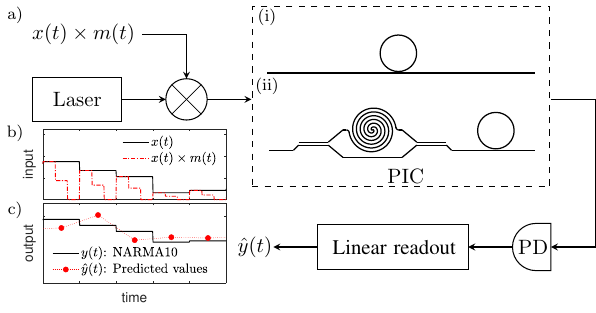}
\caption{\label{fig:FIG2} Photonic reservoir computing. a) a continuous-wave laser is modulated by a masked input signal and detected using a photodetector (PD) after passing through a photonic integrated circuit (PIC). The photodetected signal is processed at the output layer using an electronic linear readout layer. b) A typical input sequence $x(t)$ and masked signal $x(t)\times m(t)$ for a mask sequence $M=(1,0.5,0)$. c) Typical desired output signal $y(t)$ and predicted values ($\hat{y}(t)$). Note the equal symbol duration for the input $x(t)$ and the predicted output $\hat{y}(t)$ signals.}
\end{figure}

Figure \ref{fig:FIG2}a shows a schematic of the proposed reservoir computer.
We denote the input sequence to the reservoir by $X\left[n\right]$, where $n$ is an integer index. The input sequence is multiplied by a mask sequence of length $\mathcal{M} $ denoted by $M\left[n\right]$ (with \textcolor{black}{$M\left[n\right] \in [0,1] $} and again $n$ an integer index),
and is then modulated onto a continuous wave laser by either phase modulation (PM) or amplitude modulation (AM). 
For an input sequence $X\left[n\right]$ and a mask sequence $M\left[n\right]$, the input signal to the modulator is $x(t) \times m(t)$ with $t$ the time and $x\left(t\right)=X\left[n\right]$ for (n-1)$\mathcal{M} \, \tau_{p} < t < n\mathcal{M}\,  \, \tau_{p}$ for $n$ ranging from 1 to the length of input sequence.
The mask signal $m(t)$ is defined as $m\left(t\right)=M\left[n\right]$ for $(n-1) \, \tau_{p} < t < \, n \, \tau_{p}$ with $n$ ranging from 1 to $\mathcal{M}$, and is repeated by a periodicity of $\mathcal{M}\tau_p$. 
Figure \ref{fig:FIG2}b shows a typical input signal $x(t)$ and its corresponding masked signal $x(t)\times m(t)$ assuming that mask sequence is $(1,0.5,0)$. 
Masking the input signal can improve the performance of a reservoir computer, but long masks decrease the computation speed. The modulation and masking non-linearly transforms the input sequence $X\left[n\right]$ ($X\left[n\right] \in [0,1] $) into either the phase or amplitude of the mode field of the light exiting the modulator. 
We later show that for the computational tasks considered in this work, binary masking (\textcolor{black}{($M\left[n\right] \in \{0,1\} $)}) does not sacrifice the performance. 
For AM modulation, the mode amplitude at the input of PIC is $s_{\mathrm{in}}(t)=\sqrt{P_{in}\,m(t)\times x(t)}$ \cite{donati2021microring,ren2024photonic}, and for the case of PM modulation, it is $s_{\mathrm{in}}(t)=\sqrt{P_{\mathrm{in}}}\exp{\left(-{j}0.6\pi x\left(t\right)\times m\left(t\right)\right)}$ \cite{bauwens2022influence,kanno2022reservoir}. In each case, the maximum power of the modulated signal is $P_{\mathrm{in}}$. 
At low powers that is our focus in this section, complex mode amplitude at the output of the PIC $s_{\mathrm{out}}(t)$ is a linear transform of the input $s_{\mathrm{in}}(t)$, but at higher power levels the system is no longer linear. In linear regime, the system can provide an interference between the input wave and its delayed copies using a Mach-Zehnder modulator, or act as a filter using a ring resonator or other passive PIC components. After the light passes through the PIC, it is detected by a photodetector whose output is proportional to the intensity of the mode amplitude at the output ($\left|s_{\mathrm{out}}(t)\right|^2$). This is another source of nonlinearity in our reservoir system. Photodetector's output is employed in a linear readout layer which is trained to predict the desired output. Figure \ref{fig:FIG2}c shows a target signal $y(t)$ corresponding to the target sequence $Y[n]$ (which is in this example is a non-linear autoregressive moving average (NARMA) task \cite{atiya2000new}) and a typical predicted signal $\hat{y}(t)$ (corresponding to $\hat{Y}[n]$) generated by the linear readout layer.
Number of the samples used in the linear regression ($\mathcal{S}$) determines dimensionality of the output layer.  
If the detector bandwidth is large enough, it is possible to increase the dimensionality of the reservoir by oversampling the photodetector signal with frequencies that are multiples of the modulation frequency.  

We now briefly discuss the procedure to determine the parameters of the proposed optical reservoir computer. The shortest duration of the pulses $\tau_{pulse}$ that can be generated by the modulator is dictated by the available technology. Here we assume a pulse duration of $\tau_{pulse}=40 \, \mathrm{ps}$ that is readily available in silicon photonics \cite{shi2022silicon}. The nonlinear effects in a silicon MRR are usually very slow (FCA/FCD and TO effects) compared to available modulation speeds. The only exception is the Kerr effect. In spite of its speed, as shown in Fig. \ref{fig:FIG1}c this effect is weaker than FCD/FCA and TO effects and usually it is not possible to access a strong Kerr effect due to the presence of other nonlinear effects in silicon which are stronger and slower at the same time. \textcolor{black}{Reducing free-carrier lifetime in silicon by applying a bias to the nanowaveguides \cite{turner2010ultrashort} can affect the balance between different nonlinear effects and potentially improve the performance of a photonic reservoir computer \cite{giron2024effects,castro2023impact}. Another approach is based on using long mask sequences along with short input pulses (tens of picoseconds) to simultaneously increase the dimensionality of the reservoir and employ the FCD/FCA nonlinear effects  \cite{donati2021microring,donati2024time,ren2024photonic}. However, for constant pulse duration, increasing the length of the mask sequence $\mathcal{M}$ will decrease the calculation speed by a factor of $1/\mathcal{M}$. Moreover, using a long mask requires very long and usually off-chip delay lines so that the inputs at different time steps can interfere. Since long delays are not easily achievable on a PIC, it will require more complex off-chip delay lines that may compromise the stability of the reservoir \cite{donati2024time}. 
By assessing the reservoir's performance at different input power levels, we find that a ring resonator in the nonlinear regime provides almost no major improvement in the accuracy of the NARMA-10 prediction task confirming previous observations \cite{donati2021microring,donati2024time,ren2024photonic}. Hence, here we limit the input power level to very small values (\textit{e.g.} less than 10 $\mu$W) to eliminate any nonlinear effect in silicon and will show that the modulation and detection processes along with interferences present in a PIC can be effectively exploited for high-performance reservoir computation. In this approach, computation speed will no longer depend on the time scales dictated by free carrier dynamics or the thermo-optic effect.} 
There are multiple metrics for assessing the performance of a reservoir. An important property of a reservoir is its capacity to recall simple linear or nonlinear transforms of its input at previous time steps \cite{vinckier2015high}. To calculate the linear memory capacity (MC) one should train the reservoir so that its output mirrors the input at $k$ timesteps earlier for varying number of timestep delays ($k$). The input to the reservoir in this analysis is derived from a uniform distribution. MC$_k$ is now defined as
\begin{align}
&\text{MC}_k = \frac{\mathrm{cov}^2(X[n-k],Y[n])}{\mathrm{var}\left(X[n]\right)\mathrm{var}\left(Y[n]\right)}
\label{eq:MCk},
\end{align}
with $\mathrm{var}(\cdot)$ and $\mathrm{cov}(\cdot)$ the variance and covariance of the corresponding input sequences, respectively \cite{donati2021microring}. $\text{MC}_k$ expresses how well the reservoir can recall its input at $k$ timesteps earlier. The memory capacity is then calculated from
\begin{equation}
    \text{MC} = \sum_{k=1}^{k_\mathrm{max}} \mathrm{MC}_k \\,
\end{equation}
with $k_\mathrm{max}$ the largest number for which the $\mathrm{MC}_k$ is calculated.
To evaluate the potential of a reservoir computer for performing nonlinear computations, the reservoir is trained to predict a target signal which is a nonlinear transform of the input. The normalized mean squared error (NMSE) is then defined as:
\begin{equation}
\text{NMSE} = \frac{\sum{ \left( Y[n] - \hat{Y}[n] \right)^2}}{\sum \left( Y[n] - \bar{Y} \right)^2}
\label{eq:NMSE},
\end{equation}
where \( Y[n] \) is the target sequence, \( \hat{Y}[n] \) is the predicted sequence, \( \bar{Y} \) is the mean of the target sequence. 

\begin{figure}
\centering
\includegraphics[width=0.95\columnwidth]{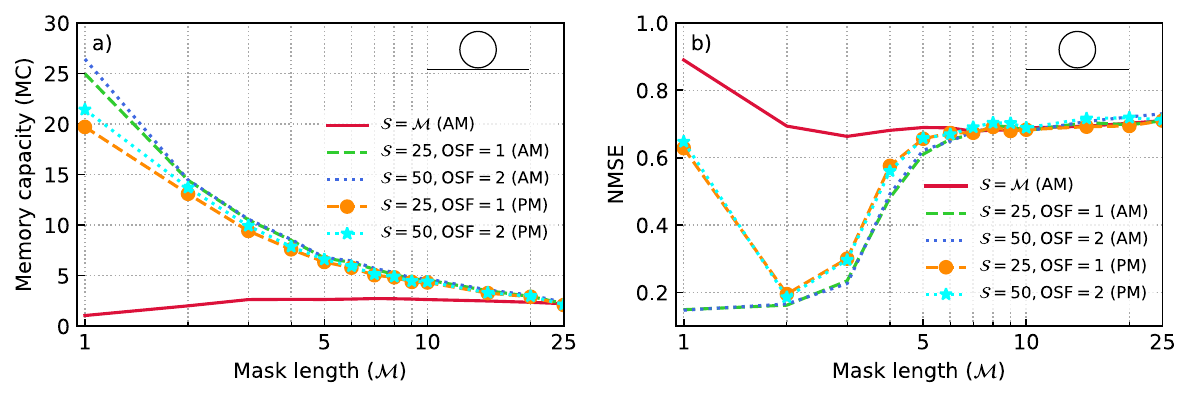}
\caption{\label{fig:FIG3} Impact of mask length on the performance of a photonic reservoir based on a single all-pass MRR. Memory capacity (MC) (a) and NMSE for the NARMA-10 prediction task (b) for different mask lengths $\mathcal{M}$,  output layer dimensionality $\mathcal{S}$, and modulation scheme (amplitude modulation (AM) or phase modulation (PM)). For the $\mathcal{M}=\mathcal{S}$ case, both the nonlinear and memory functions are totally provided by the optical components. In other cases, the memory function is partially provided by the electronic linear readout layer which significantly improves the reservoir's performance.}
\end{figure}

We investigate the impact of the mask length ($\mathcal{M}$), the detector's oversampling factor (OSF), and the number of the photodetected samples used in linear regression ($\mathcal{S}$), on the performance of the reservoir for linear memory and NARMA10 prediction tasks. For these analyses, we consider a simple PIC composed of an all-pass ring resonator. The mask sequence is a pseudorandom vector of varying length $1\leq \mathcal{M}\leq25$ with its element drawn from a uniform distribution $\mathcal{U}(0,1)$. In different analyses of this section, the mask sequence for each value of $\mathcal{M}$ is kept constant and we use a particle swarm optimization to find a combination of normalized detuning ($0\leq\delta_\lambda/\mathrm{FWHM}_\lambda\leq4$), and $0<2Q_{loaded}/Q_{int}<2$ which yields either a high MC for the linear memory task or a low NMSE for the NARMA10 task. We perform these calculations for different reservoir configurations. In all our simulations, we include the impact of the detector noise by considering a signal to noise ratio of $\mathrm{SNR}=40 \mathrm{\ dB}$ at the detector. In Fig. \ref{fig:FIG3}a and Fig. \ref{fig:FIG3}b we demonstrate the MC for the linear memory task and NMSE for the NARMA10 task, respectively. First, dimensionality of the input to the linear readout layer equals to the mask length ($\mathcal{S}=\mathcal{M}$). It is observed that increasing $\mathcal{M}$ up to two or three will improve the performance of the reservoir for both tasks due to increasing the input dimensionality. But any further increase does not provide any major improvement. 
It is possible to employ a longer mask (\textit{e.g.} $\mathcal{M}=25$) along with an external feedback loops \cite{donati2021microring,donati2024time}, or serially coupled ring resonators \cite{ren2024photonic} to achieve a higher memory capacity and consequently a better NMSE for NARMA10 task, but this comes at the cost of lowering the computation speed and more complex photonic integrated circuit. 

Another reservoir configuration can be based on storing data from previous steps in the linear readout layer as well and using it along with current data to perform the required computations. In other words, one can keep the dimensionality of the input to the linear readout layer constant (\textit{e.g.} $\mathcal{S}=25,\textrm{or}\,50$) and use masks with shorter lengths (\textit{e.g.} $\mathcal{M}=5$) so that the linear readout layer itself can complement the memory function of the PIC. \textcolor{black}{In this cooperative-memory scheme, the combination of a modulator, PIC, and the detector provides the nonlinear interaction of the nodes that are less separated in time. This is later complemented by the linear interaction of the output nodes stored in linear readout layer which can be more distant in time. The reservoir configuration with a short mask length $\mathcal{M}$ requires no different hardware than the reservoir with ($\mathcal{S}=\mathcal{M}=25$), provided that the output dimensionality is kept identical ($\mathcal{S}=25$). However, it can offer a significantly higher computation speed because it uses a shorter mask.} More importantly, it greatly reduces the stringent requirements on the delay lines and facilitates a practical realization. Although the data stored in the regression unit is only linearly transformed to calculate the final output of the RC, we find that storing the previous outputs significantly enhances the performance. Figure \ref{fig:FIG3} shows the MC and NMSE for the second reservoir configuration with $\mathcal{S}=25$ for both AM and PM modulation of the input data. For AM modulation, increasing $\mathcal{M}$ decreases the performance for both linear memory and the NARMA10 tasks. As expected, the memory capacity which is mainly provided by the regression unit is much larger than the first RC configuration. The higher memory capacity is also reflected in the substantially better performance for the  NARMA10 task. The situation is almost similar for the PM and the performance of the reservoir shows a similar trend as the AM. \textcolor{black}{In the case of PM, the memory capacity remains significant —though slightly smaller than the AM case— because the interferences occurring in the PIC practically convert phase modulation into amplitude variations which can then be detected by the intensity detector.} The other difference is that in the case of no mask ($\mathcal{M}=1$), NMSE for the PM is much larger than the AM. 
As a final configuration for the RC, we consider the case where the detector signal is over-sampled by a factor of two compared to the modulation rate ($\mathrm{OSF}=2$). Furthermore, we double the dimensionality of the linear readout layer  ($\mathcal{S}=50$). Our computations as summarized in Fig. \ref{fig:FIG3} show that in this case, the memory capacity is slightly improved, but the performance of the RC for the NARMA10 task remains unchanged. 

\subsection{MZI-MRR Configuration}\label{ResComConfig}
\begin{figure}
\centering
\includegraphics[width=1\columnwidth]{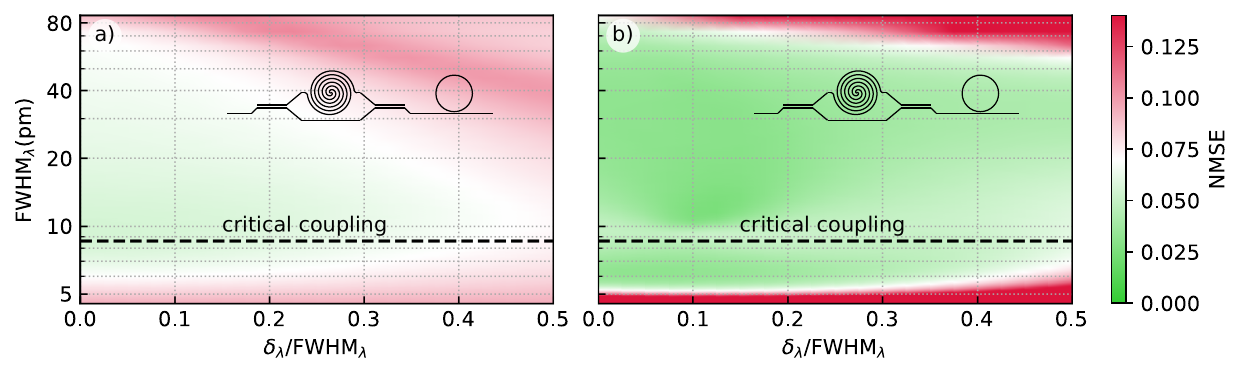}
\caption{\label{fig:FIG4} NMSE for the NARMA-10 prediction task, when the input is amplitude modulated (a) or phase modulated (b), for different values of normalized detuning and resonator linewidth.}
\end{figure}

\textcolor{black}{In the previous section we showed that the linear photonic reservoir based on a single MRR, despite its significantly simpler structure, achieves NMSE comparable to an MRR with external feedback in the NARMA-10 prediction task, provided that the reservoir's parameters are optimally selected (\textit{e.g.} $\mathcal{M}=1, \, \mathcal{S}=25$). The reservoir with a single MRR achieves such performance at considerably higher speeds than MRRs in nonlinear regime \cite{donati2021microring} because it employs no mask sequence while keeping the output layer complexity identical ($\mathcal{S}=25$). Both of these photonic reservoirs perform better than a shift register that contains the input data which achieves a minimum NMSE of 0.4 \cite{appeltant2011information}. } We can further enhance the performance of our photonic reservoir by increasing nonlinear interaction between the different nodes of the reservoir. Mach-Zehnder interferometers can provide any linear transformation \cite{reck1994experimental,miller2013self,clements2016optimal,carolan2015universal} on their inputs. Here we propose using an unbalanced MZI in conjunction with an all-pass ring resonator (Fig. \ref{fig:FIG2}a (ii)). By adjusting the delay difference between the arms of the MZI and the ratio of its input and output couplers, this configuration allows for tuning the interference between any two input nodes that are separated by a known time interval. An MZI with a large delay difference between its arms enables the interaction of input nodes that are far apart. On the other hand, an MRR whose memory capacity is related to its photon lifetime allows interaction of the nodes that are less separated in time. We now investigate how the linear MZI-MRR-based reservoir computer performs in various computational tasks starting with the NARMA-10 prediction task. For this purpose, we consider symmetric 3-dB input and output couplers at the MZI's ends to maximize the interaction of the nodes. Delay difference in MZM arms is set to  $\tau_{\textrm{delay}}=9\mathcal{M}\tau_\textrm{p}$. For mask lengths $1\leq\mathcal{M}\leq5$ the delay difference is $0.4 \textrm{ ns}\leq\tau_{\mathrm{delay}}<2 \textrm{ ns}$ which can be implemented on chip considering current available technologies \cite{ding2022high}. To simplify the practical implementation of the masking process we assume a binary mask sequence with $M[n] \in \{0,1\} $ and examine all possible binary mask sequences to find the best performance. For each possible binary mask sequence, we try to find the optimum linewidth of the cavity ($\textrm{FWHM}_\lambda$) and its normalized detuning ($\delta_\lambda/\textrm{FWHM}_\lambda$) by running a particle swarm optimization (PSO). We explore the design space by running the PSO algorithm five times with randomly generated initial particle swarms. Each run was allowed a maximum of 400 iterations. In this analysis, the output layer dimensionality is $\mathcal{S}=50$ and we keep the ring radius and the waveguide properties as before. We do these calculations for both AM and PM modulations. For AM modulation with a modulation index of 1, the best NMSE found is once the mask sequence is $M=(1,0)$ which results in $\mathrm{NMSE}=0.052$. For PM modulation, the best NMSE is less than half of the AM case ($\mathrm{NMSE}=0.024$) with the mask sequence $M=(1,0,0)$. Figure \ref{fig:FIG4} shows the calculated NMSE as a function of cavity linewidth and detuning for AM (Fig.\ref{fig:FIG4}a) and PM (Fig. \ref{fig:FIG4}b) modulations when their specific optimized mask sequence are employed. The reservoir's performance is best in a region around the critical coupling condition and slight or no detuning. In an experiment, the detuning can be controlled by adjusting the temperature of the chip or using a tunable laser \cite{padmaraju2012thermal,xie2020thermally}. The resonator linewidth, on the other hand, is controlled either by the gap between the bus waveguide and the ring resonator or electrically using a more sophisticated coupler design \cite{strain2015tunable}. 

We now investigate how MZI-MRR performs for other computing purposes. Similar analyses are performed for the Santa-Fe and Mackey-Glass prediction tasks \cite{ren2024photonic,donati2021microring}. Here we keep the MZM parameters exactly the same as before and try different modulations, masks, detunings, and coupling conditions of the MRR to find a suitable combination. In this scenario, the same PIC will handle different computational tasks with adjustments made only to its operating point. 

For the Mackey-Glass prediction task \cite{ren2024photonic,donati2021microring} and in the presence of noise (SNR=40 dB), the reservoir performs well for AM and PM modulation even when no mask sequence is applied. For AM modulation (modulation index of 1), we find NMSE=0.0035, and for PM modulation NMSE=0.002. For this task, the computation speed is the highest and is equal to the modulation speed (25 GHz).

For the Santa-Fe prediction task \cite{ren2024photonic,donati2021microring}, AM modulation and a binary mask of length 4 ($M=(1,1,0,1)$) we find $\mathrm{NMSE}=0.06$. This can be improved to $\mathrm{NMSE}=0.044$ when the modulation index is $0.5$ and the mask is ($M=(0,1,1,0)$). Fig. S1a shows the NMSE of Santa-Fe prediction tasks for different normalized detunings and coupling conditions. For PM modulation, the best performance we find is when the mask is $M=(1,0,1,1)$ that gives $\mathrm{NMSE}=0.053$(Fig. S1b). It is expected to achieve better performance if one tunes other parameters of the reservoirs as well. For example, a NMSE of 0.026 can be achieved with a mask $M=(1,0,1,1,1)$ if the sampling rate at the detector is doubled, without changing any other parameter of the reservoir. 
Table \ref{tab:RecentReservoirs} summarizes the performance of some of the most recent MRR-based photonic RCs. In all of these studies, the modulation and detection rates are 25 GHz.  Without sacrificing the accuracy, the computation speed in this work is improved due to the use of a short mask and the proposed co-operating memory configuration. In case of NARMA-10 prediction task, both the computation speed and the accuracy are significantly improved.  For the Mackey-Glass and Santa-Fe tasks, the reservoir is highly accurate and performs comparable to the previous proposals.      
\begin{table}[h]
    \centering
    \begin{tabular}{ccccccc}
         \multirow{2}{*}{Structure}& \multicolumn{2}{c}{NARMA-10} & \multicolumn{2}{c}{Mackey-Glass} & \multicolumn{2}{c}{Santa-Fe} \\  
                             & NMSE       & Rate (GHz)\textsuperscript{*}      & NMSE       & Rate (GHZ)\textsuperscript{*}      & NMSE       & Rate (GHz)\textsuperscript{*}      \\ \hline
        MRR \cite{ren2024photonic} & 0.534      & 1     & 0.0137      & 1     & 0.038      & 1     \\ 
        
         SCMRRs\cite{ren2024photonic}\textsuperscript{**}  & 0.156      & 1     & 0.0083      & 1     & 0.018      & 1     \\ \hline
         MRR+Feedback \cite{donati2021microring}  & 0.187       & 1     & 0.0014      & 1     & 0.02      & 1     \\ \hline
        MRR (AM) \textsuperscript{***}& 0.15      & 25     &       &      &       &     \\ 
        MRR (PM) \textsuperscript{***}& 0.2& 12.5     &       & & & \\ 
        MZI-MRR (AM) \textsuperscript{***}& 0.052 & 12.5 & 0.0035 & 25     & 0.044      & 6.25     \\ 
        MZI-MRR (PM) \textsuperscript{***}& 0.024 & 8.33 & 0.002      & 25  & 0.026      & 5     \\
    \bottomrule
    \end{tabular}
    \caption{Recent MRR-based photonic reservoir computers}
    \label{tab:RecentReservoirs}
    {\textsuperscript{*}\footnotesize{Computation speed assuming that the modulation and detection speed is 25 GHz}}\\{\textsuperscript{**}\footnotesize{Different PICs are used for different tasks.}}\\{\textsuperscript{***}\footnotesize{This work}}

\end{table}

\section{Optical signal equalization}\label{sec4} 
This section studies the signal equalization capabilities of the MZI-MRR structure shown in Fig. \ref{fig:FIG2} a) ii) in a numerically simulated short-reach fiber-optic transmission \textcolor{black}{\cite{kuhl2024parallel}} and compared to the feed-forward equalizer (FFE) as an established digital signal processing. It is parameterized based on recently published suggestions for the upcoming 800G standard, in which using wavelength-division multiplexing (WDM) in the O-band with \SI{112}{Gbaud} pulse-amplitude modulation (PAM) with four levels \textcolor{black}{\cite{argyris2019pam}} is proposed. In this scenario, the RC has to mostly compensate for the accumulated chromatic dispersion of the optical fiber and chirp from an electro-absorption-modulated laser (EML).

For this simulation in the O-band, a zero dispersion wavelength (ZDW) of \SI{1310}{nm} is assumed with a dispersion slope of \SI{0.09}{ps^2/nm}. With a channel separation of \SI{400}{GHz} as proposed in \cite{kuhl2024parallel}, the impact of crosstalk is negligible. Therefore, we simulate a single channel that is most impaired by chromatic dispersion and modulation chirp to investigate the worst-case scenario. For this, a pseudo-random binary sequence (PRBS) with $2^{18}$ bits is modulated using PAM-$4$. This digital signal is shaped using a raised cosine function with a roll-off factor of 0.05 before being converted into an optical signal using an EML at \SI{112}{Gbaud} with a linewidth enhancement factor $\alpha_c=-0.5$ and a \SI{3}{dB} bandwidth limitation of $\SI{50}{GHz}$ using a Butterworth filter. Before and after the signal is filtered using a Gaussian filter with a cut-off frequency of \SI{400}{GHz} to emulate the effect of (de)multiplexing. Next, the signal is coupled to the reservoir considering \SI{3}{dB} coupling losses. The output of the reservoir is amplified by a semiconductor optical amplifier (SOA) with a noise figure $n_f = \SI{6}{dB}$ to compare it to conventional signal processing based on feed-forward equalization at the same received optical power (ROP). Detection is implemented using a non-ideal photodiode with a cut-off frequency of \SI{80}{GHz}. Furthermore, a dark current of \SI{2e-8}{A}, shot and thermal noise are added to the received signal before it is resampled to \SI{224}{GSa/s}. The readout, consisting of an identical number of coefficients as the FFE, is trained to reconstruct the transmitted symbols using linear regression with Tikhonov regularization with $\alpha_{R}=1$ using singular value decomposition on 2048 training symbols. 

The reservoir's parameters are optimized using random search with 1000 iterations at an accumulated dispersion of \SI{-16}{ps/nm} and an ROP of \SI{-8}{dBm}. The normalized detuning is drawn from $d_d \sim \mathcal{N}(0, 10)$, while the MZI delay is drawn with a discrete uniform probability distribution containing half increments up to 10 of the symbol duration of $\approx \SI{4.4}{ps}$. The coupling factor for each branch is drawn from $\mathcal{U}(0,1)$ and the Q-factor from $\mathcal{U}(0.025,.975)\times Q_{int}$, ranging from strong over-coupling to extreme under-coupling. The radius of the MRR is kept constant at \SI{10}{\micro m}. The best configuration identified in this process is shown in supplementary Tab. \textcolor{black}{S2}.

\begin{figure}
\centering
\includegraphics[width=0.6\columnwidth]{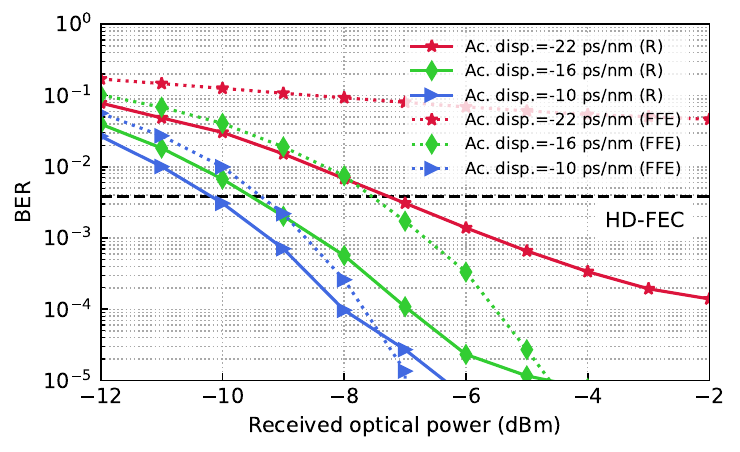}
\caption{\label{fig:FIG5} Photonic signal equalization using MZI-MRR. Bit error rate (BER) \textit{vs}. received optical power (ROP) for different accumulated dispersion when either a photonic reservoir (R) or an FFE is employed for equalization. Parameters of the photonic reservoir are optimized for an accumulated dispersion of \SI{-16}{ps/nm} and an ROP of \SI{-8}{dBm}.}
\end{figure}

As shown in Fig. \ref{fig:FIG5}, the RC is able to reduce the BER compared to an FFE of identical complexity \textcolor{black}{(with 25 first order coefficients)} as the readout. Even for higher or lower amounts of dispersion, a clear advantage of using the RC can be seen, especially for ROPs above \SI{-8}{dBm} where the system is not as severely affected by noise. Even though the RC was optimized for a single amount of accumulated dispersion, it is able to compensate \SI{-20}{ps/nm} significantly better than the FFE. At lower amounts of accumulated dispersion, only a small improvement can be seen at lower ROPs. Considering the 800G LR-4 proposal, this improvement translates to an increase in transmission reach of \SI{15}{km}, since, depending on ROP, up to \SI{-5.5}{ps/nm} of additional accumulated dispersion can be tolerated before reaching the HD-FEC limit. 

\section{Conclusions}
In this study, we find that a PIC composed of MRRs and delay elements enables high-speed computation if employed in the linear regime. Not limited by the free carrier lifetime or other comparably slow nonlinear processes, these reservoirs rely on modulation \ detection nonlinearity. Our studies also show that the memory capacity of the reservoir is effectively enhanced once the electronic linear readout layer keeps data of the previous time steps. The MZI-MRR structure proposed in this work, along with the cooperative reservoir configuration, enables high-accuracy and high-speed computation for all different tasks explored in this research. The computation speed in this configuration is higher than in previous studies without employing any different hardware, but by tailoring the masking, modulation, and linear response of the PIC. Here, we find that either of AM or PM modulation schemes can perform better according to the computational task being investigated. A more detailed study of the modulation/detection schemes may reveal ways to improve the nonlinear function of the reservoir and the computation performance. Since the reservoir computer proposed in this research does not depend on free-carrier nonlinearity, increasing the computation speed even up to 110 GHz is easily accessible considering the state-of-the-art integrated optical modulators. This will simplify the design of the PIC as well because delay lines with shorter lengths and ring resonators with lower Q-factors will be needed. Furthermore, the MZI-MRR configuration proposed here can be extended to an array with elements that are different in their delay lines and MRRs, in order to perform more complex computations. In such a scenario, the electronic storage of the data along with the rich and diverse interference of the optical signals in the PIC will contribute to high-speed and accurate photonic computation.  
\section{Acknowledgment}

{
This work was supported in part by the German Research Foundation (Deutsche Forschungsgemeinschaft, DFG);  project number: 498410117. 

}

\bibliography{sn-bibliography}

\end{document}